\documentclass[a4paper]{article}
\usepackage{vmargin}
\usepackage{amssymb,latexsym}
\usepackage{amsmath, amsthm}
\usepackage{enumerate}

\DeclareFontFamily{OT1}{rsfs}{} \DeclareFontShape{OT1}{rsfs}{m}{n}{
<-7> rsfs5 <7-10> rsfs7 <10-> rsfs10}{}
\DeclareMathAlphabet{\mycal}{OT1}{rsfs}{m}{n}

\newtheorem{Theorem}{Theorem}
\newtheorem{Proposition}{Proposition}

\newtheorem{Corollary}{Corollary}
\newtheorem{Remark}{Remark}

\newcommand{\rd}{{\rm d}} 

\newcommand{\lie}{{\mathcal{L}}} 
\newcommand{\vol}{\hat{\epsilon}}

\newcommand{\Lie}{\lie}   
\newcommand{\contr}{{ {}^{{\hphantom{.}}_{{}_{\scriptstyle{\lrcorner}}}} }}
\DeclareMathOperator*{\const}{const}

\DeclareMathOperator*{\mdiv}{div}
\DeclareMathOperator*{\hdiv}{\hat{\mdiv}}
\newcommand{\IH}{\triangle}
\newcommand{\sli}{S} 

\newcommand{\D}{\mathbb{D}}
\newcommand{\hq}{\hat{q}}
\newcommand{\hw}{\hat{\omega}}
\newcommand{\cF}{\mathcal{F}}
\newcommand{\hF}{\hat{\mathcal{F}}}
\newcommand{\hst}{\hat{*}}
\newcommand{\hD}{\hat{D}}

\newcommand{\litem}{\item \vspace{-.1in}} 

\begin{document}

\begin{center} 
\begin{minipage}[c]{\textwidth} 
\renewcommand*{\thempfootnote}{\arabic{mpfootnote}} 
\renewcommand*{\footnoterule}{} 
\title{\bf Spacetimes foliated by Killing horizons}
\author{Tomasz Pawlowski${}^{1,3}$, Jerzy Lewandowski${}^{1,3,4}$, 
        Jacek Jezierski${}^{2,3}$}
\date{}
\maketitle

\footnotetext[1]{Instytut Fizyki Teoretycznej, Uniwersytet Warszawski,
  ul. Ho\.{z}a 69, 00-681 Warsaw, Poland} 
\footnotetext[2]{Katedra Metod Matematycznych Fizyki, Uniwersytet Warszawski,
  ul. Ho\.{z}a 74, 00-682 Warsaw, Poland}
\footnotetext[3]{Max Planck Institut f\"ur Gravitationsphysik, 
  Albert Einstein Institut, 14476 Golm, Germany} 
\footnotetext[4]{Physics Department, 104 Davey, Penn State, University 
  Park, PA 16802, USA} 
\end{minipage} 
\end{center} 
\smallskip

\begin{abstract}
  It seems to be expected, that  a horizon of a quasi-local type,
  like a Killing or an isolated horizon, by analogy with a globally defined
  event horizon, should be unique in some open neighborhood in the spacetime,
  provided the vacuum Einstein or the Einstein-Maxwell equations are satisfied.
  The aim of our paper is to verify whether that intuition is correct.
  If one can extend a so called Kundt metric, in such a way that its null,
  shear-free surfaces have spherical spacetime sections, the resulting 
  spacetime is foliated by so called non-expanding horizons. The obstacle  is
  Kundt's constraint induced at the surfaces by the Einstein or the 
  Einstein-Maxwell equations, and the requirement that a solution be globally 
  defined on the sphere. We derived a transformation (reflection) that creates 
  a solution to  Kundt's constraint out of data defining an extremal isolated 
  horizon. Using that transformation, we derived a class of exact solutions 
  to the Einstein or Einstein-Maxwell equations of very special properties.
  Each spacetime we construct is foliated by a family of the Killing
  horizons.  Moreover, it admits another, transversal Killing horizon.
  The intrinsic and extrinsic geometry of the transversal Killing horizon  
  coincides with the one defined on the event horizon of the extremal 
  Kerr-Newman solution. However, the Killing horizon in our example admits 
  yet another Killing vector tangent to and null at it. The geometries of the 
  leaves are given by the reflection.
\end{abstract}


\section{Introduction}

In the standard black-hole theory an event horizon is defined as a
boundary of certain distinguished region of spacetime. On the
other hand, quasi-local definitions of horizons are known which
lead to a local black-hole physics and geometry. Typically, a
horizon is a cylinder formed by 2-surfaces diffeomorphic with the
2-sphere. In the stationary black hole  case, the cylinder is
expected to be a null surface. If the intrinsic geometry of the
surface is preserved by a tangent null flow, then the cylinder is
called a non-expanding, shear-free horizon. If both, the intrinsic
and extrinsic geometries are preserved by a tangent null flow,
then the cylinder becomes an isolated horizon (see \cite{ABDFKLW,ABL}
for details). In a very special case, there is a Killing vector
field defined in a neighborhood of the cylinder and tangent to its
null generators. Then we deal with a Killing horizon. It seems to
be expected, that  a horizon of this quasi-local type, by analogy
with the globally defined event horizon, should be unique in some
open neighborhood in the spacetime. The aim of our paper is to
verify whether that intuition is correct. We would like to prove
or disprove the existence of a solution of vacuum Einstein's or
the Einstein-Maxwell equations foliated by non-expanding,
shear-free horizons. A metric tensor of this type necessarily
belongs to the Kundt's class \cite{exact}. The structure of
Einstein's and the Einstein-Maxwell equations in this class is
discussed in \cite{exact} and a large family of exact solutions is
known. The new issue is the quasi-global assumption, that the
null, non-expanding, shear-free surfaces foliating a given Kundt
solution have space-like spherical sections.

We approach this problem in terms of the non-expanding horizons
geometry. Vacuum Einstein's equations imposed on a metric tensor
which admits a foliation by non-expanding, shear-free horizons
induce on each horizon  certain constraint equations on the
intrinsic geometry and the rotation 1-form potential.  We
construct a transformation that maps the constraints into another
set of equations satisfied by the intrinsic geometry and the
rotation 1-form potential of an extremal isolated horizon
contained in a vacuum spacetime. The inverse transform applied to
the intrinsic geometry and the rotation 1-form potential defined
on the extremal Kerr event horizon provides a solution to our
constraints! The vacuum constraints and the transformation are
generalized to the Einstein-Maxwell case. Whereas the nature of
that transformation is somewhat mysterious in the general case
considered here, it becomes clear in the case of the examples we
derive later. Next, we extend appropriately every solution to the
vacuum (electrovac) constraint equations  into a solution to the
vacuum Einstein (Einstein-Maxwell) equations. The derived class of
solutions admits a two dimensional Lie algebra of the Killing
vector fields generated by $K_0, K_1$ which satisfy
$[K_0,K_1]=K_0$. For every value of $u$, the vector field $uK_0
-K_1$ defines a bifurcated Killing horizon  (that is it is tangent
to a pair of  Killing horizons which share a spherical slice). All
the bifurcated Killing horizons share a single Killing horizon
tangent to $K_0$.  In other words, the resulting spacetime is
foliated by the Killing horizons, and admits one more Killing
horizon transversal to the leaves of the foliation. The
transversal Killing horizon equipped with the Killing vector $K_0$
is an extremal vacuum (electrovac) isolated horizon. Its intrinsic
geometry, the rotation 1-form potential, and the electromagnetic
field are related to the data defined on each lief of the
foliation {\it exactly by the transformation discussed above}.
Finally, we derive an explicit form of the solution corresponding
to the extremal Kerr-Newman horizons.

The constraint equations considered as equations of certain data
defined on a manifold diffeomorphic with a  2-sphere  are
interesting by themselves.  According to the results of
\cite{extremal}, the only axi-symmetric solutions are those given
by the Kerr-Newman extremal horizon and the transform. It is not
known if there are any other solutions. Some partial results are
enclosed in Appendix.

In conclusion, our results contradict even the belief that  a
Killing horizon should be unique in some spacetime neighborhood.
However our examples are quasi local and their geodesic extension
should be understood.

\section{Non-expanding horizons}\label{sec:def} 

We recall in this
section the definition of the non-expanding horizons and
properties relevant in this paper \cite{ABL}. A non-expanding horizon
is defined  to be a null 3-surface $\IH$ contained in a
4-dimensional spacetime of the Lorentz signature, such that:
\begin{enumerate}[ (i)]
  \item $\IH$ is an embedding of $S\times R$, where the manifolds
        $S$ and $R$ are, respectively, the 2-sphere and the 1-dimensional
        interval, 
  \item for every $s\in S$, the embedding maps$\{s\}\times R$ into a null 
        curve, 
  \item every null vector field tangent to $\IH$ is non-expanding.
\end{enumerate}

It follows from the Raychaudhuri equation, that
if the stress energy tensor satisfies at every point of $\IH$
the non-negativity condition
\begin{equation}\label{eq:Tll}
  T_{\mu\nu}\ell^\mu\ell^\nu\ge 0,
\end{equation}
where $\ell$ is a null vector tangent to $\IH$, then $\IH$ is also
shear free. It means that the pullback $q$ of the spacetime
metric $g$ onto $\IH$  is transversal to $\ell$ and  Lie dragged
by $\ell$,
\begin{subequations}\label{eq:lq}\begin{align}
  \ell \contr q \ &= \ 0,  &
  \Lie_\ell q\ =\ 0. \tag{\ref{eq:lq}}
\end{align}\end{subequations}
It follows, that the spacetime covariant derivative $\nabla$
reduces naturally to a covariant derivative $\D$ defined in the
tangent bundle  $T\IH$, that is for every two vector fields $X,Y$
tangent to $\IH$, so is $\nabla_XY$. It also follows that the null
direction tangent to $\IH$ is covariantly constant, hence for
every null vector field $\ell$ tangent to $\IH$,
\begin{equation}
  \D\ell\ =\ \omega\otimes \ell,
\end{equation}
where $\omega$ is a differential 1-form defined on $\IH$, called
the rotation 1-form potential. We can always choose $\ell$ to be
geodesic, that is such that $\nabla_\ell\ell=0$. Then necessarily
\cite{ABL}
\begin{subequations}\label{eq:lomega}\begin{align}
  \ell \contr \omega\ &=\ 0,  &
  \Lie_\ell \omega\ &=\ 0.\tag{\ref{eq:lomega}}
\end{align}\end{subequations}

Note, that every non-orientable $S$ admits a double covering by
the orientable one. Therefore we assume in this paper that $S$ is
orientable and fix an orientation in $S$. Let as fix an
orientation and a time orientation in $M$. An orientation in $S$
is adjusted in the following way: Let $\ell$ be a future oriented
null vector tangent to $\IH$ at $x$, and $n$ be another future
oriented null vector at $x$. A frame $(X,Y)$ tangent to $S$ at
$p(x)$, where where $p:\IH \rightarrow S$, is the natural
projection,  has a positive orientation whenever the  the
orientation of the frame $(X,Y,n,\ell)$ is positive.

 If the non-expanding horizon is contained in an Einstein-Maxwell
vacuum, then the electromagnetic field tensor $F$ defines on $\IH$
yet another transversal to (and Lie dragged by) $\ell$ object,
namely the pullback ${\cF}$ of the self dual part
$\frac{1}{2}(F-\star F)$ of $F$ onto $\IH$,
\begin{subequations}\label{eq:lF}\begin{align}
  \ell \contr {\cF} \ &= \ 0,  &
  \Lie_\ell {\cF} \ &= \ 0.  \tag{\ref{eq:lF}}
\end{align}\end{subequations}

It follows from (\ref{eq:lq},\ref{eq:lomega},\ref{eq:lF}) that, there are 
defined on $S$: a metric tensor $\hq$, a 1-form $\hw$ and a complex
valued differential 2-form $\hF$, such that
\begin{subequations}\label{eq:proj_data}\begin{align}
  q \ &= \ p^*\hq,  & 
  \omega \ &= \ p^*\hw,  & 
  {\cF} \ &= \ p^*\hF,  \tag{\ref{eq:proj_data}}
\end{align}\end{subequations}
 We call $(\hq, \hw, \hF)$ respectively: {\it a
projected 2-metric tensor, a projected rotation 1-form potential,
and a projected electromagnetic field 2-form induced on $S$ by $(\IH,\ell)$.}
The 2-form $\hF$ can be represented by a {\it projective  complex 
electromagnetic scalar $\Phi_1$ respectively defined on $S$ by}:
\begin{equation}
  \hF\ =:\ i\Phi_1{\vol} 
\end{equation}
where ${\vol}$ is the volume 2-form on $S$ defined by the metric
tensor $\hq$ and the orientation.   

Note, that given $\ell$ as above can be always replaced by
$\ell'=f\ell$, where $f$ is a function such that $\ell^\mu
f_{,\mu}=0$. The metric tensor  $\hq$ and the complex valued
differential 2-form ${\cF}$ defined on $S$ are independent of
that choice, whereas $\omega'=\omega + d\ln f$. We will see below,
however, that in the case a 1-dimensional family of non-expanding
horizons foliating a spacetime, that freedom will be restricted
to $f=\const$.

In the example of spacetime  found in this paper we encounter a
pair of non-expanding horizons which share a sphere. If this is
the case, the structures induced on $S$ are related to each other.
Indeed, suppose $\IH'$ is another non-expanding horizon in $M$
such that $\IH\cap \IH'=\tilde{S}$ where $\tilde{S}$ is
diffeomorphic to a sphere. Let $\ell'$ be a null vector field
tangent to $\IH'$ such that
\begin{equation}\label{eq:bifur_ell}
  {\ell^\mu\ell'_\mu}|_{\tilde{S}}=\ -1.
\end{equation}
The spheres of the null geodesics $S$ and $S'$ can be naturally
identified. Then, the corresponding projective metric tensors
$\hq$ and $\hq'$ and the projective electromagnetic field
2-forms $\hF, \hF'$ just coincide. The orientations of $S$ corresponding 
to $\IH$ and $\IH'$ are opposite to each other. The pullbacks 
$\omega_{(\tilde{S})}$ and $\omega'_{(\tilde{S})}$ of the rotation 1-form 
potentials $\omega$ and $\omega'$ to $\tilde{S}$ are related in the 
following way
\begin{equation}
\omega'_{(\tilde{S})} =
\left(\ell_\mu\nabla\ell'^\mu\right)_{(\tilde{S})} =
-\left(\ell'_\mu\nabla\ell^\mu\right)_{(\tilde{S})}=
-\omega_{(\tilde{S})}.
\end{equation}
In summary, we have
\begin{subequations}\label{eq:bifur_data}\begin{align}
  \hq'\ &=\ \hq,  &
  \hw'\ &=\ -\hw,  &
  \Phi'_1\ &=\ -\Phi_1,  &
  {\rm orientation}'\ &=\ -{\rm orientation}. \tag{\ref{eq:bifur_data}}
\end{align}\end{subequations}

\section{Foliations by non-expanding horizons}\label{sec:foliation}

We consider in this work a spacetime $(M,g)$  foliated by
non-expanding horizons. Our considerations are quasi local, in the
sense that we  assume that
\begin{enumerate}[ (i)]
  \item $\hphantom{.}$ \vspace{-.15in}
    \begin{equation}\label{eq:prod}
      M\ =\ S\times R\times  R,
    \end{equation}
    where $S$ is a manifold diffeomorphic with the 2-sphere, and $R$
    is an interval,
    \label{NEHfolDef_c1}
  \item for every  $u'\in R$, the 3-surface
    $S\times R\times \{u'\}$ is a non-expanding horizon,
    \label{NEHfolDef_c2}
  \item every curve $\{s\}\times R\times \{u'\}$
    such that $s\in S$ and $u'\in R$ is null.
    \label{NEHfolDef_c3}
\end{enumerate}

Every spacetime foliated by the non-expanding horizons can be
introduced (quasi-locally) in this way.  Let $u$ be a real valued
function  defined on $M$ such that for every $u'\in R$, $u$ is
constant on the corresponding non-expanding horizon $S\times R\times \{u'\}$
and
\begin{equation}
  du\ \neq\ 0,
\end{equation}
for every $x\in M$. A function $u$ of those properties can be given by
any coordinate defined in $R$. For every value $u_0$ of the function $u$,
the non-expanding horizon $S\times R\times \{u'\}$ on which $u$ takes that
value henceforth will be  denoted by $\IH_{u_0}$ (often we will drop the
suffix $0$ at $u$). The function $u$ provides in $M$ a  null vector field
$\ell$,
\begin{equation}\label{eq:du}
  \ell^\mu\ :=\ -g^{\mu\nu}u_{,\nu}
\end{equation}
geodesic,
\begin{equation}
  \nabla_\ell \ell\ =\ 0,
\end{equation}
and tangent at every $x\in M$ to $\IH_{u(x)}$. Therefore,  for
every non-expanding horizon $\IH_{u}$ the results of the previous
section apply to the vector field $\ell$ defined in \eqref{eq:du}.
In conclusion, for every lief $\IH_{u}$ of the foliation, $S$ is
equipped with: the metric tensor $\hq$, the
projected rotation 1-form potential $\hw$ and the
projected self-dual electromagnetic field $\hF$ as it was
explained in the previous section. The structure
$(\hq,\,\hw,\,\hF)$ arbitrarily depends on on the lief  $\IH_{u}$.

There is still some freedom in the definition of the vector field
$\ell$ on $M$. The function $u$ can be replaced by $\tilde{u}=
h(u)$ where $h$ is an arbitrary function whose gradient nowhere
vanishes. That transformation amounts to the rescaling of $\ell$ by
a factor $h'(u)$  constant on every lief $\IH_u$. The transformation
leaves  $(\hq,\,\hw,\,\hF)$ invariant.
Therefore the structure $(\hq,\,\hw,\,\hF)$ is uniquely defined on $S$ 
and depends only on the lief $\IH_u$ and the spacetime metric tensor $g$.

\section{Constraints}\label{sec:constr} 

Imposing the Einstein-Maxwell equations on the spacetime foliated by 
non-expanding horizons implies interesting constraints on the structures 
$(\hq,\hw,{\cF})$ defined on $S$. We formulate and discuss
the constraints in this section. The derivation will be
presented in the Section \ref{sec:proof} in which we analyze the full set of
the Einstein-Maxwell equations imposed on a spacetime foliated by
the non-expanding horizons. Let us begin with the  Einstein vacuum
case, that is $F=0$ in $M$.

\begin{Proposition}\label{thm:constr_vac} Suppose $(M,g)$ is a 4-dimensional 
  spacetime foliated by non-expanding horizons in the meaning of the conditions
  (\ref{NEHfolDef_c1})-(\ref{NEHfolDef_c3}) of Section \ref{sec:foliation}; 
  suppose  $g$ satisfies the vacuum Einstein equations. Then, the following 
  constraint is satisfied on the manifold $S$ for every lief $\IH_u$,
  \begin{equation}\label{eq:constr_vac}
    \hD_B\hw_{A} + \hD_A\hw_{B} 
    - 2\hw_{A}\hw_{B}
    \ +\ \hat{R}_{AB}\ =\ 0,
  \end{equation}
  where we denote by $\hat{D}$ and $\hat{R}$ the covariant
  derivative and the Ricci tensor, respectively, of the 2-metric
  tensor $\hq$ defined on $S$.
\end{Proposition}

In the presence of the electromagnetic field $F$, a constraint
implied by the Maxwell equations will be added (and the first
constraint will be modified by the non-zero $F$). It is convenient
to  express the  Maxwell constraint by the complex structure on
$S$ compatible with the orientation of $S$ distinguished in
Section \ref{sec:def}. Let $(z,\bar{z})$ be a local complex coordinate
system in $S$ such that the orientation of the coframe $(\Re dz,
\Im dz)$ is consistent with the orientation of $S$. We decompose
every differential 1-form $k=k_z dz+ k_{\bar{z}}d\bar{z}$ defined
on $S$ in the following way,
\begin{subequations}\label{eq:compl_dec1f}\begin{align}
  k^{(1,0)} \ &= \ k_zdz,  &
  k^{(0,1)} \ &= \ k_{\bar z}d\bar{z},  \tag{\ref{eq:compl_dec1f}}
\end{align}\end{subequations}
and in particular for every complex-valued function $f$ defined on
$S$ we define
\begin{subequations}\label{eq:compl_dec0f}\begin{align}
  \partial f\ &=\ df^{(1,0)},  & 
  \bar{\partial} f\ &=\ df^{(0,1)}.  \tag{\ref{eq:compl_dec0f}}
\end{align}\end{subequations}

\begin{Proposition}\label{thm:constr_evac} Suppose $(M,g)$ is a 4-dimensional 
  spacetime foliated by non-expanding horizons in the meaning the conditions 
  (\ref{NEHfolDef_c1})-(\ref{NEHfolDef_c3}) of Section \ref{sec:foliation}; 
  suppose $F$ is an electromagnetic vector field in $M$ such that  $(g, F)$ 
  satisfy the vacuum Einstein-Maxwell equations. Then, the following 
  constraints are satisfied on the manifold $S$ for every leaf $\IH_u$,
  \begin{subequations}\label{eq:constr_evac}\begin{align}
    \label{eq:constr_evac_qw} 
      \hD_B\hw_{A} + \hD_A\hw_{B} 
      - 2\hw_{A}\omega_{B} \ +\ \hat{R}_{AB}
      \ - 2\kappa_0|\Phi_1|^2\hq_{AB}\ &=\ 0, \\
    \label{eq:constr_evac_F}
      (\partial  - 2\hw^{(1,0)})\Phi_1\ &=\ 0.
  \end{align}\end{subequations}
\end{Proposition}
The rotation 1-form $\hw$ can be decomposed on the sphere in the following way:
\begin{equation}\label{eq:omega_dec}
  \hw \ = \ \hst\rd U + \rd\ln B  
\end{equation}
where $U,B$ are real functions. 
Then constraint \eqref{eq:constr_evac_F} can be easily integrated giving the 
following form of $\Phi_1$ as a function of $(U,B)$,
\begin{subequations}\label{eq:int_phi1}\begin{align}
  \Phi_1\ &=\ E_0 B^2 e^{2iU} \ ,  &
  E_0\ &=\ \const \ .              \tag{\ref{eq:int_phi1}}
\end{align}\end{subequations}
This equation was investigated in more details in \cite{extremal}.

Obviously Proposition \ref{thm:constr_vac} is a special case of Proposition 
\ref{thm:constr_evac}. The proof of Proposition \ref{thm:constr_evac} follows 
in a straight forward way from the Einstein-Maxwell equations discussed in 
the Section \ref{sec:proof}.

The constraints considerably restrict possible 2-metric tensors
$\hq$, the rotation 1-form potential $\hw$ and electromagnetic field   
induced on $S$. Their strength consist in fact that solutions have to be 
defined globally on $S$.

On the other hand, we will also  show in Section \ref{sec:completion}, that 
for every single solution of the constrains \eqref{eq:constr_evac} there 
exists an electrovac $(M,g,F)$ foliated by non-expanding horizons.

\section{A transform providing solutions to the constraints} 

There is a remarkable mathematical relation between the constraints
\eqref{eq:constr_vac} and the constraints satisfied by the geometry of a
vacuum extremal isolated horizon \cite{extremal}. The relation generalizes
to the constraints \eqref{eq:constr_evac} and the constraints
satisfied by the geometry and  the electromagnetic field
on an extremal electrovacuum  isolated horizon \cite{extremal}.
(By `mathematical', we mean that the constraints are not
the same, but there is a transformation that maps solutions
of one set of the constraints into another)
Let us recall, that an extremal isolated horizon $(\IH',\ell')$ in a
non-expanding horizon $\IH'$ equipped with a tangent null vector
field such that
\begin{equation}
  [\Lie_{\ell'},{\D'}]\ =\ 0,\ {\rm and}\ {\D'}_{\ell'} {\ell'}\ =\ 0.
\end{equation}
At $\IH'$, the  vacuum Einstein equations induce the following constraint
equations on the metric tensor $\hq'$ and the rotation 1-form
potential $\hw'$ projected onto the manifold $S$ \cite{extremal} 
(we are still using the notation of Section \ref{sec:def})
\begin{equation}\label{eq:EIH-vac}
  \hD'_B\hw'_A + \hD'_A\hw'_B 
  + 2\hw'_A\hw'_B 
  - \hat{R}'_{AB} \ = \ 0
\end{equation}
Comparing the equations above with the vacuum constraints
\eqref{eq:constr_vac} one can easily see the relation:

\begin{Theorem}\label{thm:trans_vac}
  The following map
  \begin{subequations}\label{eq:trans_vac}\begin{align}
    \hw'\ &\mapsto\ \hw\,=\,\-\hw',  &
    \hq'\ &\mapsto\ \hq\,=\,\hq'  \tag{\ref{eq:trans_vac}}
  \end{align}\end{subequations}
  is a bijection of the set of solution to the vacuum extremal isolated
  horizon constraints (\ref{eq:EIH-vac}) onto the set of solutions to the
  constraints (\ref{eq:constr_vac}). In particular, the projected metric tensor
  $\hq'$ and the projected rotation 1-form potential  $\hw'$ induced on the 
  event horizon in the extremal Kerr spacetime are mapped by the transformation
  (\ref{eq:trans_vac}) into a solution to the constraints 
  (\ref{eq:constr_vac}).    
\end{Theorem}

In the presence of the electromagnetic field $F'$ at an extremal isolated
horizon $\IH'$, the Einstein-Maxwell equations and an assumption that  the
electromagnetic field is Lie dragged by the vector field $\ell'$
(see \cite{extremal} for the details) amount to the following constraint
equations
\begin{subequations}\label{eq:EIH-elvac}\begin{align}
  \hD'_B\hw'_A + \hD'_A\hw'_B 
    + 2\hw'_A\hw'_B 
    - \hat{R}'_{AB} + 2\kappa_0|\Phi'_1|^2\hq_{AB} \ &= \ 0 \\
  \left( \bar{\partial} + 2\hw^{(0,1)} \right) \Phi'_1 
    \ &= \ 0
\end{align}\end{subequations}
A comparison of the equations above with the electrovac
constraints \eqref{eq:constr_evac} shows that Theorem 1
generalizes to the following:

\begin{Theorem}\label{thm:trans_evac} 
  The map (\ref{eq:trans_vac}) accompanied by
  \begin{equation}\label{eq:trans_evac}
    \Phi'_1\ \mapsto\ \Phi_1\,=\,\overline{\Phi'_1}
  \end{equation}
  is a bijection of the set of solutions to the extremal electrovac isolated
  horizon constraints (\ref{eq:EIH-elvac}) onto the set of solutions to the
  constraints (\ref{eq:constr_evac}). In particular, the projected metric 
  tensor  $\hq'$, the projected rotation 1-form potential  $\hw'$ and the 
  projected electromagnetic field $\hF'$  induced on the event horizon in the 
  extremal Kerr-Newman spacetime 
  are mapped by the transformation (\ref{eq:trans_vac},\ref{eq:trans_evac}) 
  into a solution to the constraints (\ref{eq:constr_evac}).    
\end{Theorem}

The relevance of Theorem 1 and Theorem 2 above consists in establishing the 
existence of non-trivial, globally defined on $S$ solutions to the 
vacuum constraints \eqref{eq:constr_vac} and, respectively, the electrovac
constraints  \eqref{eq:constr_evac}.    

\begin{Remark}
  Applied to the constraints (\ref{eq:constr_evac}),
  the transformation (\ref{eq:trans_vac},\ref{eq:trans_evac}) is equivalent to
  the transformation (\ref{eq:bifur_data}) mapping into
  each other  data corresponding to intersecting two non-expanding
  horizons. That observation may be considered as indication of the
  possible  existence of a non-expanding horizon transversal to one
  of the horizons $\IH_u$. This is exactly what happens in the case
  of the class of solutions constructed in the next section.
\end{Remark}

\section{Proof of Proposition \ref{thm:constr_vac},\ref{thm:constr_evac}}
\label{sec:proof}

Every spacetime $(M, g)$ foliated by non-expanding horizons in the sense of  
the conditions \eqref{NEHfolDef_c1}-\eqref{NEHfolDef_c3} of Section 
\ref{sec:foliation} can be represented by $M=S\times R\times R$ and the 
following metric tensor
\begin{equation}\label{eq:Kundt_g}
  g\ =\ \hq - 2du\left(dv + \hat{W} +Hdu \right)
\end{equation}
where: 
\begin{enumerate}[a)]
  \litem the function $v$ (respectively, $u$) is a
        parametrization of the first (second) factor $R$ of the product
        \eqref{eq:prod} extended naturally to $S\times R\times R$,
  \litem $\hq$ is a metric tensor defined on $S$ and depending on value
        of $u$, naturally lifted to the product $S\times R\times R$,
  \litem $\hat{W}$  ( $H$ ) is a differential 1-form (a function)
        defined on $S$ depending on values of $u$ and $v$ and extended
        naturally to the product $S\times R\times R$. 
\end{enumerate}
Let $z,\bar{z}$ are any (local) coordinate system defined on $S$ and extended
naturally to some domain in $S\times R\times R$. In terms of the
coordinate system $(z, \bar{z}, v, u)$ the eq. \eqref{eq:Kundt_g} reads,
\begin{subequations}\label{eq:metr4Pv}\begin{align}
  g\ &=\ 2P^{-2}\rd z \rd\bar z - 2 \rd u \left(\rd v
      + W \rd z+\overline{W}\rd\bar z+ H \rd u \right)   &
  P_{,v}\ &=\ 0,         \tag{\ref{eq:metr4Pv}}
\end{align}\end{subequations}
where $P$ is a real-valued and $W$ is a complex-valued function.
The constancy surfaces of $u$ are non-expanding horizons.

The converse is also true: given a  metric tensor \eqref{eq:metr4Pv},
a surface $\IH_{u_0}=S\times R\times \{u_0\}$ is a non-expanding horizon for
every value of $u_0$ taken by the functions $v,u$ respectively.

We express now all the structures defined in 
Sections \ref{sec:def}, \ref{sec:foliation} by the components of the metric 
tensor above.
To begin with, the function $u$ is the same as in Section
\ref{sec:foliation}, and the complex valued function $z$ is the same
as $z$ in Section \ref{sec:constr}.
In this coordinate system the vector field $\ell$ \eqref{eq:du} is
\begin{equation}
  \ell^{\mu}\partial_{\mu} \ = \ \partial_v.
\end{equation}
The structures $\hq$ and $\hw$ introduced in $S$ for
every value taken by the function $u$ are:
\begin{subequations}\label{eq:Kundt_qw}\begin{align}
  \hq \ &= \ 2P^{-2}\rd z \rd\bar{z},  &
  \hw \ &= \ \frac{1}{2}\left(W_{,v}dz + \bar{W}_{,v}d\bar{z}\right).
    \tag{\ref{eq:Kundt_qw}}
\end{align}\end{subequations}

Note that Eq. \eqref{eq:lomega} is equivalent to
\begin{equation}
  W_{,vv}\ =\ 0.\label{eq:Wvv}
\end{equation}

Every electromagnetic field $F$ defined on $M$ which satisfies the
conditions \eqref{eq:lF} can be written as
\begin{subequations}\label{eq:F_frame}\begin{align}
  F\ &=\ \Phi_1 (e^4 \wedge e^3 + e^2 \wedge e^1) +  \Phi_2 e^2 \wedge e^3
      + \overline{\Phi_1} (e^4 \wedge e^3 - e^2 \wedge e^1) 
      + \overline{\Phi_2} e^1\wedge e^3   &
  \Phi_{1,v}\ &=\ 0,  \tag{\ref{eq:F_frame}}
\end{align}\end{subequations}
where $(e^1, e^2, e^3, e^4)$ is the coframe dual to the following null
frame
\begin{subequations}\label{eq:frame}\begin{align}
  e_1 = \bar{e}_2\ &=\ P\partial_{z},  &
  e_3\ &=\ \partial_u + P^2(\bar{W}\partial_{z}+W\partial_{\bar{z}})
        - (H+P^2W\bar{W})\partial_v,  &
  e_4\ &=\ \partial_v.  \tag{\ref{eq:frame}}  
\end{align}\end{subequations}
Then, $\Phi_1$ is the same function as the one introduced in Section 
\ref{sec:def} and the complex valued differential 2-form $\hF$ defined in
the same section on $S$ is
\begin{equation}
  \hF\ =\ -\frac{\Phi_1}{P^2}d\bar{z}\wedge dz.
\end{equation}

In the construction of we have already took into account a part of
the Einstein-Maxwell equations. Therefore, our metric tensor $g$
(\ref{eq:metr4Pv}, \ref{eq:Wvv}) and the electromagnetic field
$F$ \eqref{eq:F_frame} already satisfy
\begin{subequations}\begin{align}
  R_{44} = \kappa_0T_{44} \ &= \ 0 \ 
         = \ R_{41} = \kappa_0T_{41} \ = \ R_{42} = \kappa_0T_{42},\\
  e_1\contr e_4\contr d(F -i*F)
    \ &= \ 0 \ = \ e_2\contr e_4\contr d(F -i*F)
\end{align}\end{subequations}

Every metric tensor given by (\ref{eq:metr4Pv}, \ref{eq:Wvv})
belongs to the Kundt's class. We apply below the discussion of the
structure of the Einstein-Maxwell equations for this class which
can be found in \cite{exact}.
\bigskip

We turn now to the proof of Proposition \ref{thm:constr_vac} and Proposition
\ref{thm:constr_evac}. Consider the following Einstein equations

\begin{align}
    (P^2W_{,v})_{,z} - \frac{1}{2}P^2(W_{,v})^2= R_{11}\ 
     &= \ \kappa_0 T_{11}= 0   \\
  \label{eq:Ricci12}
    2P^2(\ln(P))_{,z\bar{z}} + \frac{1}{2}P^2(\bar{W}_{,vz}+W_{,v\bar{z}}
                             - W_{,v}\bar{W}_{,v})=R_{12}\ &=\ \kappa_0 T_{12}
                             = 2\kappa_0 \Phi_1\overline{\Phi_1} \\
\end{align}
where in each of the lines above, the first and the third equality
is an identity. Using the relations between $P$ and the 2-metric
tensor $\hq$ and between $W_{,v}$ and $\hw$ (see
\ref{eq:Kundt_qw}) one easily recognizes Eq. \eqref{eq:constr_evac_qw} and, in
the vacuum case,  Eq. \eqref{eq:constr_vac}.

Finally, one of the Maxwell equations reads
\begin{equation}\label{eq:Maxw1z}
  e_3\contr e_4\contr e_1\contr d(F-i\star F)
  = \frac{1}{P}\left( \Phi_{1,z}\ -\ W_{,v}\Phi_1 \right)\ =\ 0
\end{equation}
where the first equality is an identity.   This is exactly Eq.
\eqref{eq:constr_evac_F}. This completes the proof.

\section{A class of solutions to the Einstein-Maxwell equations}
\label{sec:completion}

Let us assume in this section, that we are given
$W_{,v}(z,\bar{z},u)$, $P(z,\bar{z},u)$ and $\Phi_1(z,\bar{z},u)$
such that the constraints (\ref{eq:constr_evac_qw},\ref{eq:constr_evac_F}) 
are satisfied for every sphere $u=\const,\, v=\const$. We turn now to
the issue of the existence of solutions to the full set of the
Einstein-Maxwell equations.

The second derivative with respect to $v$ of the function $H$ is
given by the $(\mu, \nu)=(3,4)$ component of the Einstein-Maxwell
equations, namely
\begin{equation}\label{eq:Ricci34}
  H_{,vv}\ =\ 2\kappa_0|\Phi_1|^2 + \hdiv\hw - 2\hq^{AB}\hw_A\hw_B 
  \ =\ 2\kappa_0 |\Phi_1|^2 +  \frac{1}{2}P^2(\bar{W}_{,vz}
  + W_{,v\bar{z}}  -2W_{,v}\bar{W}_{,v}) 
\end{equation}
where $\hdiv\hw:=\hq^{AB}\hD_A\hw_B$ and $\hq^{AB}$ is the two-dimensional
inverse metric.

Similarly, the $v$ derivative of the remaining component $\Phi_2$ is
involved in  $e_3\contr e_2\contr e_4\contr d(F-i\star F)$
and the corresponding Maxwell equation gives
\begin{equation}\label{eq:Maxw2v}
  \Phi_{2,v}\ =\ P(\Phi_{1.\bar{z}}+\bar{W}_{,v}\Phi_1).
\end{equation}
In particular, it follows from (\ref{eq:Ricci34},\ref{eq:Maxw2v})
that $H_{,vvv}=0$ and $\Phi_{2,vv}=0$. Hence the most general
form of the functions $W, H, \Phi_2$ is
\begin{subequations}\begin{align}
  \label{eq:dec_W}
    W(z,\bar{z},u,v)\ &=\ W_{,v}(z,\bar{z},u)v
                       + W_0(z,\bar{z},u) \\
  \begin{split}
    H(z,\bar{z},u,v)\ &=\ \left[ \kappa_0 |\Phi_1|^2 
                                 + \frac{1}{4}P^2(\bar{W}_{,vz}
                                 + W_{,v\bar{z}} - 2W_{,v}\bar{W}_{,v})
                          \right] v^2 \\
                      &+\ G_0(z,\bar{z},u)v
                       + H_0(z,\bar{z},u) 
  \end{split} \\
  \Phi_2(z,\bar{z},u,v)\
     &=\ P(\Phi_{1,\bar{z}}+\bar{W}_{,v}\Phi_1)v
      + \Phi_2^0(z,\bar{z},u)
\end{align}\end{subequations}

The remaining Einstein-Maxwell equations amount to the following
system of equations on $W_0,G_0,H_0, P_{,u}, W_{,vu}$ \cite{exact}
(the first two equations are explicitly written down in
\cite{exact}, but to propose a solution we needed also the last
equation below, therefore the first two were included for the
completeness)

\begin{subequations}\label{eq:completion}\begin{align}
  \begin{split}
    2\kappa_0 \Phi_1 \Phi_2^0\
      &=\ P(P^2W^0)_{,z\bar{z}}
       + P\left[ (\ln(P))_{,u} - \frac{1}{2}P^2\bar{W}_{,z}^0
                 - \frac{1}{2}P^2W_{\bar{z}}^0 \right]_{,z} \\
      &+\ \frac{1}{2}P \left[ (P^2W_{,v})_{,z}\bar{W}^0
                              + (P^2\bar{W}_{,v})_{z}W^0 \right]
       - 2P_{,\bar{z}}(P^2W^0)_{,z} \\
      &+\ PG_{,z}^0
       + \frac{1}{2}P^3 \left[ (W^0\bar{W}_{,v})_{,z}
                               - (W^0W_{,v})_{,\bar{z}} \right]
       + P_{,u} - \frac{1}{2}PW_{,vu}
  \end{split} \\
  P\Phi_{2,z}^0\ &=\ P_{,z}\Phi_2^0
    + \Phi_{1,u} - 2(\ln(P))_{,u}\Phi_1
    + P^2 \left[ (\bar{W}^0\Phi_1)_{,z}
                 + (W^0\Phi_1)_{,\bar{z}} \right] \\
  \begin{split}
    2\kappa_0 \Phi_2^0 \bar{\Phi}_2^0\
      &=\ 2P^2H_{,z\bar{z}}^0
       + P^2\left[ (\bar{W}_{,v}H^0)_{,z}
       + (W_{,v}H^0)_{,\bar{z}} \right] \\
      &+\ 2P^2s_{,z\bar{z}}^0
       + P^2\left[ (\bar{W}_{,v}s^0)_{,z}
                   + (W_{,v}s^0)_{,\bar{z}} \right]
       - 2(P^2\bar{W}^0)_{,\bar{z}}(P^2W^0)_{,z} \\
      &-\ 2 \left[ P^2(\mu_{,z}^0\bar{W}^0
                   + \mu_{,\bar{z}}^0W^0) + \mu_{,u}^0
                   + \mu^0\left( G^0 + P^2(W^0\bar{W}_{,v}+\bar{W}^0W_{,v}) 
                                 + \mu^0 \right)
            \right]
  \end{split}
\end{align}\end{subequations}
where:
\begin{subequations}\label{eq:H0_aux}\begin{align}
  s^0\ &=\ P^2W^0\bar{W}^0 &
  \mu^0\ &=\ \frac{1}{2}P^2(\bar{W}_{,z}^0+W_{,\bar{z}}^0)
          - (\ln(P))_{,u}  \tag{\ref{eq:H0_aux}}
\end{align}\end{subequations}

It is easy to see that the following example defines a solution
of \eqref{eq:completion}
\begin{equation}\label{eq:prez}
  P_{,u} = W_{,vu}= \Phi_{1,u} \ = \ 0 \ = \ W^0 = G^0 = H^0 = \Phi_2^0.
\end{equation}
Therefore, the following has been shown:

\begin{Theorem}\label{thm:existence}
  For every 2-metric tensor $\hq$, a differential
  1-form $\hw$ and a complex valued function $\Phi_1$ all
  defined globally on $S$ and solving the constraint equations
  (\ref{eq:constr_evac_qw},\ref{eq:constr_evac_F}), the metric tensor $g$ and
  electromagnetic field $F$ defined  by \eqref{eq:prez} are a solution of the
  Einstein-Maxwell equations. If $\Phi_1=0$, then all the resulting
  electromagnetic field $F=0$.
\end{Theorem}

\begin{Corollary} Combining Theorem \ref{thm:existence} with Theorem 
  \ref{thm:trans_vac} and Theorem \ref{thm:trans_evac} we established 
  the existence  of solutions to vacuum Einstein's equations foliated 
  by non-expanding horizons, and solutions to the vacuum Einstein-Maxwell 
  equations foliated by non-expanding horizons.
\end{Corollary}

We discuss now the structure of the spacetime given by Theorem
\ref{thm:existence} and by a solution $(\hq,\hw,\Phi_1)$
of the constraint equations (\ref{eq:constr_evac_qw},\ref{eq:constr_evac_F}).

The metric tensor $g$ has two Killing vector fields, namely
\begin{equation}\label{eq:spc_Kil}
  K_1\ =\ \partial_u \quad \text{ and } \quad  
  K_2\ =\ u\partial_u - v\partial_v.
\end{equation}
For every value of $u_0$, the corresponding lief $\IH_{u_0}$ is
tangent to the Killing vector field $K_2 - u_{0}K_1$. Moreover,
$K_2 - u_{0}K_1$ is null on $\IH_{u_0}$. Therefore, all the leaves
are the Killing horizons. Every $(\IH_{u_0}, K_2 - u_{0}K_1)$ is
non-extremal in the sense, that on $\IH_{u_0}$,
\begin{equation}
  \nabla_{(K_2 - u_{0}K_1)} (K_2 - u_{0}K_1) 
  \ =\ -(K_2 - u_{0}K_1)\ \neq\ 0.
\end{equation}

The Killing vector $K_1$ is transversal to every lief $\IH_{u_0}$.
The 1-dimensional group of isometries it generates maps one lief
into another. The spacetime pseudo-norm of $K_1$ equals
\begin{equation}
  g_{uu}\ =\ -2H\ =\ - \left[ 2\kappa_0 |\Phi_1|^2 
                              + \frac{1}{2}P^2(\bar{W}_{,vz}
                              + W_{,v\bar{z}} - 2W_{,v}\bar{W}_{,v})
                       \right] v^2\ .
\end{equation}
Therefore, $K_1$ becomes null on the surface $v=0$. Let as denote
that surface by $\IH^o$.  The surface $\IH^o$ is null itself and
tangent to $K_1$.  Hence $\IH^o$ it is yet another Killing
horizon. It has certain quite special property: there are two
distinct null Killing vectors tangent to $\IH^o$. Indeed, the
other one is the vector field $K_2$. On $\IH$ they satisfy
\begin{subequations}\label{eq:Kil_geod}\begin{align}
  \nabla_{K_1}K_1\ &=\ 0,  &
  \nabla_{K_2} K_2\ &=\ K_2.  \tag{\ref{eq:Kil_geod}}
\end{align}\end{subequations}
Therefore $(\IH, K_1)$ is extremal whereas $(\IH, K_2)$ is
non-extremal.

Consider finally the projected metric, projected rotation 1-form
potential and the projected electromagnetic field defined on the 
sphere $S$ by $(\IH^o, K_1)$ (see Section \ref{sec:def}). 
Denote them by $\hq^o, \hw^o$ and $\hF^o$ respectively. Because the 
horizons $\IH^o$ and $\IH_{u_0}$ share a sphere, and
\begin{equation}
  g(K_1,\ell)\ =\ g_{uv}\ =\ -1,
\end{equation}
the structure $(\hq^o, \hw^o, \hF^o)$ is related  to the structure 
$(\hq, \hw, \hF)$
\begin{subequations}\label{eq:Kil_bifur_data}\begin{align}
  \hq^o \ &= \ \hq,  &
  \hw^o \ &= \ -\hw,   &
  \Phi^o_1 \ &= \ \bar{\Phi}_1.  \tag{\ref{eq:Kil_bifur_data}}
\end{align}\end{subequations}
This is exactly the transformation (\ref{eq:trans_vac}, \ref{eq:trans_evac}).

\section{The axi-symmetric example}

In the previous section we constructed a class of solutions to the
Einstein-Maxwell equations labeled by solutions to the constraints 
\eqref{eq:constr_evac}. We complete now 
the task of the construction of an explicit electrovac example by deriving
the axi-symmetric solutions to the constraints.\footnote{
  The general solution $(\hq,\hw,\Phi_1)$ of the constraint equations
  \eqref{eq:constr_evac} is not known.  See \cite{extremal} and Appendix 
  for discussion. The general axi-symmetric solution was derived in 
  \cite{extremal}. It is given via Theorem \ref{thm:trans_evac} by the 
  Kerr-Newman solution. Since the eq. (131a) in \cite{extremal}
  contains a misprint made in the process of rewriting the results,
  we outline  the derivation in this section.}
In the axi-symmetric case  we can choose spherical coordinates
$(\theta,\varphi)$ such that the functions $P,W_v,\Phi_1$ are
independent on $\varphi$. We replace the coordinate $\theta$  by a
function $z(\theta)$ such that:
\begin{subequations}\label{eq:def_x}\begin{align}
  P^2 dx\ &=\ d\theta &
  x\ &\in\ \left[ -\frac{A}{4\pi},\frac{A}{4\pi} \right], &
  A\ &=\ \int_{S} \vol
    \tag{\ref{eq:def_x}}
\end{align}\end{subequations}
The condition that the 2-metric tensor  $\hq$ (see \eqref{eq:Kundt_qw}) 
is  of the class $C^1$ on $S$ implies the following conditions on the 
function $P$ to be satisfied at the poles:
\begin{subequations}\label{eq:glob_cond}\begin{align}
  \lim_{x\to\pm\frac{A}{4\pi}} \frac{1}{P}\ &=\ 0  &
  \lim_{x\to\pm\frac{A}{4\pi}} \left( \frac{1}{P^2} \right)_{,x}\ &=\ \mp 2
    \tag{\ref{eq:glob_cond}}
\end{align}\end{subequations}
After applying this coordinate system and the equation \eqref{eq:int_phi1},
the constraint \eqref{eq:constr_evac_qw} (with $\hw$ decomposed by 
\eqref{eq:omega_dec}) takes the following form:
\begin{subequations}\label{eq:basic_axi}\begin{align}
  \label{eq:basic_axi_trace}
    \left( \frac{1}{PB} \right)_{,xx}
    + \frac{(P_{,x})^2}{P^3B}
    + \frac{U_x^2}{PB}
    + 2\kappa_0|E_0|^2PB^3\ &=\ 0 \\
  \label{eq:basic_axi_trless}
    \left( \frac{1}{B} \right)_{,xx} - \frac{(U_{,x})^2}{B}\ =\ 0
      \qquad\qquad
    \left( \frac{U_{,x}}{B^2} \right)_{,x}\ &=\ 0
      \tag{\ref{eq:basic_axi}b,c}
\end{align}\end{subequations}
where \eqref{eq:basic_axi_trace} is the trace of \eqref{eq:constr_evac_qw} and
\eqref{eq:basic_axi_trless} is the result of the splitting of the traceless 
part of \eqref{eq:constr_evac_qw} into the real and, respectively, imaginary 
part.

The set of solutions to the equations \eqref{eq:basic_axi} and 
to the globality conditions \eqref{eq:glob_cond} can be labeled
by  three real parameters $\alpha, A, \theta_0$ such that:
\begin{subequations}\label{eq:sol_par}\begin{align}
  \alpha   \ &\in\ [0,1],      &
  A        \ &\in\ ]0,\infty[, &
  \theta_0 \ &\in\ [0,2\pi[.   \tag{\ref{eq:sol_par}}
\end{align}\end{subequations}
And the general solution is:  
\newcommand{\etaq}{\frac{1-\alpha^2}{1+\alpha^2}}
\begin{subequations}\label{eq:axi_sol}\begin{align}
  P^2(x) \ &= \ \frac{2\pi(1+\alpha^2)}{A}\
             \frac{A^2+\etaq (4\pi x)^2}{A^2-(4\pi x)^2}, \\
  U(x) \ &= \ \pm\arctan\left( 4\pi\sqrt{\etaq}\frac{x}{A} \right), \\
  B(x) \ &= \ \left( 1+\etaq \frac{(4\pi x)^2}{A^2} \right)^{-\frac{1}{2}}, \\
  \Phi_1(x) \ &= \ e^{i\theta_0} \sqrt{\frac{2\pi}{\kappa_0}}\
                \frac{ 2\alpha A^{\frac{3}{2}}}{1+\alpha^2}\
                \frac{ \left(A^2-\etaq(4\pi x)^2\right)
                       \pm 2iA\sqrt{\etaq}(4\pi x) }
                     { \left( A^2+\etaq(4\pi x)^2 \right)^2 }
\end{align}\end{subequations}
It is easy to show \cite{extremal} that this solution corresponds via
the transformation $P\to P,\ \omega\to-\omega,\ \Phi_1\to\bar{\Phi}_1$)
to the extremal Kerr-Newman event horizon.  The case $\alpha=0,1$, 
in particular, corresponds to the extremal Kerr and the extremal 
Reissner-Nordstr\"{o}m event horizons, respectively.

In conclusion, the class of explicit examples of electrovacua foliated by 
the Killing horizons is given  by the metric tensor \eqref{eq:metr4Pv} and the 
electromagnetic field \eqref{eq:F_frame} such that:
\begin{subequations}\label{eq:axi_sol_Kil}\begin{align}
  P^2 \ &= \ \frac{2\pi(1+\alpha^2)}{A}\
             \frac{A^2+\etaq (4\pi x)^2}{A^2-(4\pi x)^2} \ ,\\
  W \ &= \ \frac{2\sqrt{2}(1-\alpha^2)^{\frac{1}{2}}A}
                {(1+\alpha^2)^{\frac{3}{2}}}\
           \frac{A^2 - (4\pi x)^2}
                {\left( A^2 + \etaq(4\pi x)^2 \right)^2}\
           \left( \pm iA \ - \ \sqrt{\etaq}(4\pi x) \right) v \ ,\\
  H \ &= \ 4\pi A \frac{ \left( A^2 + \etaq(4\pi x)^2 \right)^2
                         - 4\etaq A^2 \left( A^2 - (4\pi x)^2 \right) }
                       { (1+\alpha^2) \left( A^2 + \etaq(4\pi x)^2 \right)^3 }
         \, v^2 \ ,\\
  \Phi_1 \ &= \ e^{i\theta_0} \sqrt{\frac{2\pi}{\kappa_0}}\
                \frac{ 2\alpha A^{\frac{3}{2}}}{1+\alpha^2}\
                \frac{ \left(A^2-\etaq(4\pi x)^2\right)
                       \pm 2iA\sqrt{\etaq}(4\pi x) }
                     { \left( A^2+\etaq(4\pi x)^2 \right)^2 } \ ,\\
  \Phi_2 \ &= \ e^{i\theta_0} \sqrt{\frac{2}{\kappa_0}}\ 
                \frac{\alpha (1-\alpha^2) (8\pi A)^2 x}
                     {(1+\alpha^2)^{\frac{5}{2}}}\
                \frac{ \left( A^2 - (4\pi x)^2 \right)^{\frac{1}{2}} }
                     { \left( A^2 + \etaq(4\pi x)^2 \right)^{\frac{7}{2}} }
                \left( A \pm i\sqrt{\etaq}(4\pi x) \right)^2 v  \ .
\end{align}\end{subequations}

In particular, the vacuum solutions are the type D Kundt's solutions
expressed in \cite{exact} by:
\begin{subequations}\label{eq:type_D}\begin{align}
  P^2 \ &=\ \frac{x^2+l^2}{k((x^2-l^2)+2mx)}\ , \\
  W   \ &=\ -\frac{\sqrt{2}v}{P^2(x-il)} \ , \\
  H   \ &=\ -\left( \frac{k}{2(x^2+l^2)} 
                    + \frac{2l^2}{P^2(x^2+l^2)^2} \right)v^2 \, , 
\end{align}\end{subequations}
where in our case the parameters  $k,l,m$ are real and 
such that:
\begin{subequations}\label{eq:type_D_glob}\begin{align}
  m       \ &=\ 0               & 
  -k = 2l \ &=\ \frac{A}{2\pi}.  \tag{\ref{eq:type_D_glob}}
\end{align}\end{subequations}
Comparing our results with \cite{exact} we may also conclude, that
the vacuum solutions derived in this section provide all the 
vacuum and Petrov type $D$ Kundt's spacetimes foliated by the non-expanding
horizons.

\section*{Acknowledgments}

We would like to thank Abhay Ashtekar, Jiri Bicak, Ted Jacobson, Jorma Louko, 
and Bernd Schmidt for discussions.
This work was supported in part by the Polish Committee for Scientific Research
(KBN) under grants no: \mbox{2 P03B 073 24}, \mbox{2 P03B 127 24},
\mbox{2 P03B 130 24}, the Albert Einstein Institute of the Max Planck Society,
and Batory fellowship.

\appendix
\renewcommand{\thesection}{Appendix \Alph{section}:} 

\section{A topological constraint} 

From the geometrical point of view the definition of a non-expanding 
horizon may be  generalized to an arbitrary compact 2-manifold $S$. 
It is convenient to assume $S$ is orientable and notice that every 
non-orientable $S$ case can be obtained by dividing of an orientable 
case by a discrete group of symmetries.
Then still the definitions of the structures $(\hq,\hw,\hF)$ introduced 
in Section \ref{sec:def} apply, as well as Proposition \ref{thm:constr_vac} 
and Proposition \ref{thm:constr_evac}.
Let as consider that generalization  in this section (only) to prove
the following topological consequence of the constraints:
\begin{Theorem}\label{thm:topol}
  Suppose $(M,g,F)$ is an electrovac foliated by non-expanding horizons
  in the meaning of the conditions (\ref{NEHfolDef_c2}),(\ref{NEHfolDef_c3}) 
  of Section \ref{sec:foliation}
  and the condition (\ref{NEHfolDef_c1}) replaced by the following assumption:
  $M= S\times R\times R$, where $S$ is a compact, orientable 2-manifold
  and $R$ is a manifold diffeomorphic with an interval.
  Then $S$ is either the torus or the sphere. In the first case, the only
  solution of the constraints (\ref{eq:constr_evac}) is
  \begin{subequations}\label{eq:torus_data}\begin{align}
    \hw \ &= \ 0,  &
    \hF \ = \ 0    \tag{\ref{eq:torus_data}}
  \end{align}\end{subequations}
  and a flat metric $\hq$.
\end{Theorem}

\begin{proof}
  The trace part of \eqref{eq:constr_evac_qw} reads
  \begin{equation}\label{eq:divom}
    d\hat{\star}\hw\ +\ K{\vol} \ 
      =\ \left(\hq^{AB}\hw_A \hw_B 
      +  2\kappa_0|\Phi_1|^2\right){\vol},
  \end{equation}
  where $\hat{\star}$ and  $K:=\frac{1}{2}\hq^{AB} R_{AB}$ are,
  respectively,  the Hodge star and the Gaussian curvature of
  $(S,\hq)$, and $\hq^{AB}$ is the two-dimensional inverse
  metric.  The integral of the equation along $S$ and the
  Gauss-Bonnet theorem give the genus of $S$,
  \begin{equation}\label{eq:GB}
    2-2\mathbf{g}\ =\ \int_S\left(\hq^{AB}\hw_A \hw_B 
      + 2\kappa_0|\Phi_1|^2\right)\vol.
  \end{equation}
  It follows immediately that
  \begin{equation}
    \mathbf{g}\ \le\ 1.
  \end{equation}

  In the case of $\mathbf{g}=1$, that is when $S$ is a torus, the
  equation \eqref{eq:GB} implies that $\hw$ and $\Phi_1$ are
  identically $0$. Then, a consequence of \eqref{eq:constr_evac_qw} is that
  the metric $\hq$ on $S$ is Ricci flat.
\end{proof}

In conclusion the only compact, orientable 2-manifolds which
admits a solution of the constraints \eqref{eq:constr_evac}
are the 2-sphere and the 2-torus. It is not justified, however,
to refer to a non-expanding null 3-surface whose section
is topologically  a torus as to a `horizon', because a surface
of those properties  can be  admitted even by a flat geometry.

\section{The vacuum constraint equations}

In this section we will restrict ourselves to solutions to the constraint
equations \eqref{eq:constr_evac} with $\Phi_1 = 0$. This assumption doesn't
restrict us to vacuum spacetimes - null electromagnetic radiation is still 
allowed.

By contracting traceless part of \eqref{eq:constr_evac_qw} with $\hw^A\hw^B$ 
(where indexes of $\hw$ are raised using inverse metric $\hq^{AB}$ on $\sli$)
we obtain the following identity:
\begin{equation}
  \label{eq:trless_con}
    \hq^{AB}\hw_A \hD_B|\hw|^2 \ = \ |\hw|^2\hdiv\hw + |\hw|^4\ ,
\end{equation}
where
\begin{subequations}\label{eq:trless_con_sup}\begin{align}
  |\hw|^2  \ &=\ \hq^{AB}\hw_A\hw_B\ ,  &
  \hdiv\hw \ &=\ \hq^{AB}\hD_A\hw_B\ .  \tag{\ref{eq:trless_con_sup}}
\end{align}\end{subequations}
From this equation and \eqref{eq:divom} follows, that the following 
equality holds for each real $\beta$
\begin{equation}\label{eq:trless_par}
  \hq^{AB} \hD_A(|\hw|^{2\beta}\hw_B) \ = \ (2\beta+1)|\hw|^{2(\beta+1)}
     - (\beta+1)|\hw|^{2\beta}K\ .
\end{equation}
That finally implies one-parameter family of integral identities
\begin{equation}\label{eq:int_norm}
  \frac{2\beta+1}{\beta+1}\int_{\sli} |\hw|^{2(\beta+1)} \vol \ =\
  \int_{\sli} K |\hw|^{2\beta}  \vol \ .
\end{equation}
Suppose $\hw$ has only finite set of critical points (that is such that 
$|\hw|=0$)
which are isolated.
Then equation \eqref{eq:trless_par} for $\beta=-\frac12$ takes the form
\begin{equation}\label{eq:norm-K}
  \hq^{AB} \hD_A \frac{\hw_B}{|\hw|} \ = \ -\frac{K}{2|\hw|} \ .
\end{equation}
That equation can be integrated over $\sli$ avoiding the singularities
by surrounding the critical points by small circles and
passing to the limit (i.e. shrinking circles to critical points).
It provides the special case of the equation \eqref{eq:int_norm}
\begin{equation}\label{eq:int_norm-K}
  \int_{\sli} \frac{K}{|\hw|}\vol \ = \ 0 \ .
\end{equation}
This condition implies that $K$ must be negative on some open subset of 
$\sli$. The considerations can be summarized by the following
\begin{Theorem}\label{thm:neg_K}
  There are no solutions to constraints \eqref{eq:constr_evac} of the
  following properties:
  \begin{enumerate}[ (a)]
    \item projected electromagnetic field tensor $\hF$ vanishes,
    \litem rotation 1-form $\hw$ vanishes only at finite set of points,
    \litem $\sli$ is a sphere with non-negative Gaussian curvature.
  \end{enumerate}
\end{Theorem}
Gauss-Bonett theorem implies that $\hw \neq 0$ on an open subset. On the other
hand, solution generated by projected data of extremal Reissner-Nordstr\"{o}m 
horizon (case $\alpha=1$ in \eqref{eq:axi_sol_Kil} is an example of solutions 
for Einstein-Maxwell equations with $\hw \equiv 0$, which means that some 
arguments used above are no longer true for solutions describing fields other 
than null radiation.

\end{document}